\documentclass[twoside]{dis04}
\def\runauthor{W. Broniowski and E. Ruiz Arriola}
\def\shorttitle{Kwieci\'nski evolution of UPDs}

\usepackage{epsfig}
\usepackage{graphics}
\usepackage{graphicx}
\usepackage{amsfonts}
\usepackage{amssymb}
\usepackage{amsmath}
\usepackage{cite}
\usepackage{subfigure}

\begin{document}

\title{Kwieci\'nski evolution of unintegrated parton distributions }

\author{{\underline{Wojciech Broniowski}${}^1$} and Enrique Ruiz Arriola${}^2$}

\address{${}^1$The H. Niewodnicza\'nski Institute of Nuclear Physics,\\
Polish Academy of Sciences, PL-31342 Krak\'ow, Poland\\
${}^2$Departamento de F\'{\i}sica Moderna, Universidad de
Granada, \\ E-18071 Granada, Spain}

\maketitle

\abstracts{The Kwieci\'nski evolution of unintegrated
parton distributions (UPDs) in the transverse-coordinate space is
analyzed with the help of the Mellin transform. 
Numerical results are presented for the unintegrated pion
distributions with a simple valence-like initial condition at the low
scale, which follows from chiral large-$N_c$ quark models. 
The effect of spreading of UPDs in the transverse
momentum with the increasing scale is confirmed, with $\langle
k_\perp^2 \rangle$ growing asymptotically as $Q^2 \alpha(Q^2)$.
Formal aspects of the equations, such as the limits
of UPDs at $x \to 0$, $x \to 1$, and at low and large $b$, are 
straightforward to obtain with our method. }

This talk is based on Refs.~\cite{GKB,ERAWB} and focuses on numerical as
weel as formal 
aspects of the QCD evolution of UPDs proposed by Kwieci\'nski \cite{JK}.
Practical applications are presented by Szczurek \cite{AS}, while similar approaches are
discussed by Jung \cite{Jung} and Lonnblad \cite{LL} in these proceedings.

The definition of the leading-twist UPDs \cite{Collins} may be read-off from Fig.~1(a). The 
transverse momentum in the partonic loop, $k_\perp$, is left {\em unintegrated}
and is not constrained.
\begin{figure}[b]
\begin{center}
\subfigure{\includegraphics[angle=0,width=0.5\textwidth]{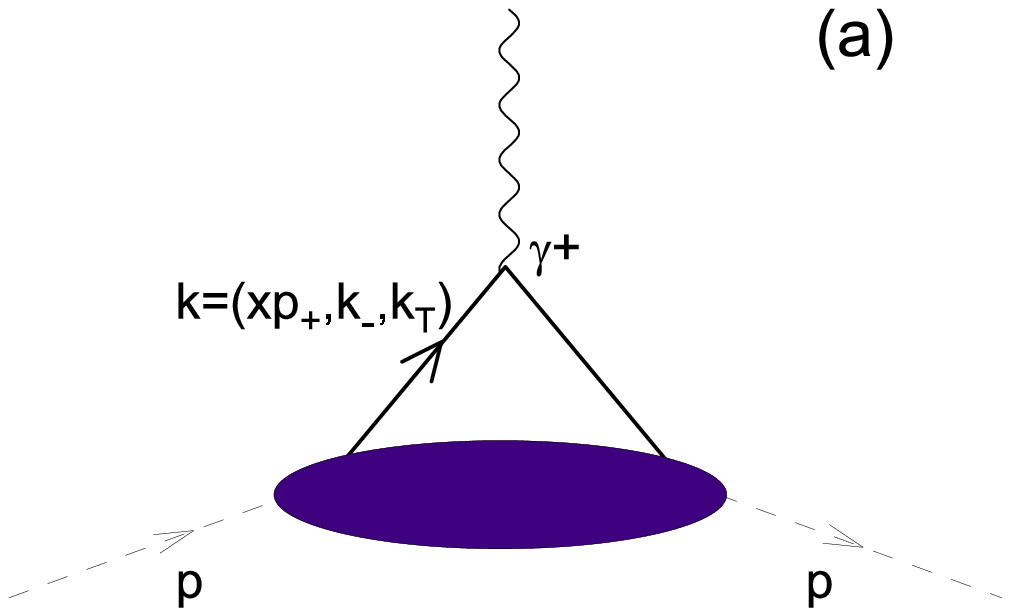}} \hfill
\subfigure{\includegraphics[angle=0,width=0.36\textwidth]{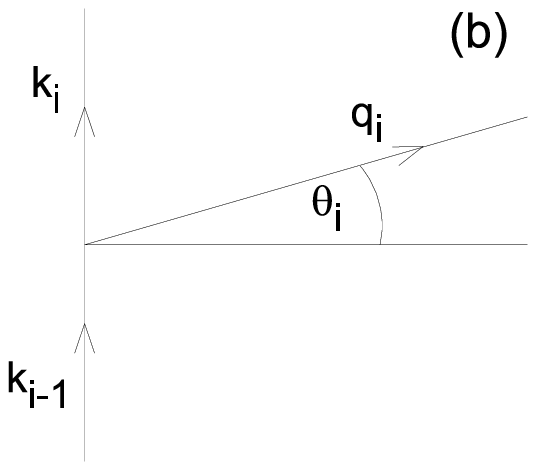}}
\end{center}
\caption{(a) Definition of the leading-twist UPD. (b) Kinematics inside 
the partonic cascade.}
\end{figure}
Kwieci\'nski's approach is based on the CCFM 
formalism with the following modifications:
\begin{itemize}
\item One-loop CCFM is used, with the angular ordering replaced with a
stronger condition on the scaled gluon momenta of the form 
$q'_{\perp,i} > q'_{\perp,i-1}$, where $q'_i\equiv q_i/(1-z_i)$,
$z_i \equiv {x_i}/x_{i-1}$, and $x_i$ denotes the fraction of the 
longitudinal momentum of the hadron carried by the parton.
\item Both singlet and non-singlet quarks are included.
\item Non-Sudakov form factor is set to unity.
\item The evolution is written in the transverse-coordinate space, $b$, 
Fourier-conjugated to the transverse momentum.
\end{itemize}
The Kwieci\'nski equations are diagonal in the $b$-space and read \cite{JK}:
\begin{eqnarray}
&&Q^2{\partial f_{\rm NS}(x,b,Q)\over \partial Q^2} = \label{Keq} \\
&&{\alpha_s(Q^2)\over 2\pi}  \int_0^1dz  \,P_{qq}(z)
\bigg [ \Theta(z-x)\,J_0((1-z)Qb)\, f_{\rm NS}\left({x\over z},b,Q\right)
- f_{\rm NS}(x,b,Q) \bigg ], \nonumber
\end{eqnarray}
with similar coupled equations for the sea quarks and gluons. 
Note that for the integrated PDs (case of $b=0$) the equations assume the familiar
DGLAP form. 

In this talk we consider the uPDF of the pion. 
The initial conditions for the evolution are taken from the chiral quark models 
\cite{ERAWB,DavArr,spectr}, which in the chiral limit provide a particularly simple, factorized form.
Moreover, these conditions are consistent with the formal requirements, such as proper
normalization, correct support, crossing symmetry, and the momentum sum rule. 
The working scale of the model, $Q_0$, is estimated using the fact that at $Q_0$ 
all the momementum is carried by the quarks, which are the only degrees
of freedom at that scale (see \cite{ERAWB} for details). The Spectral
Quark Model \cite{spectr} provides a factorized initial condition of the form
\begin{eqnarray}
q(x,b,Q_0)&=& F(b) \theta(x)\theta(1-x), \label{sqmb} \\
F(b) &=& \left (1+\frac{b M_V}{2} \right )
 \exp\left ( - \frac{M_V b}{2} \right )  \nonumber
\end{eqnarray}
for the valence quarks, with no gluons nor sea quarks at the scale $Q_0$.
Note that the functional dependence in $b$ in Eq.~(\ref{sqmb}) factorizes in Eq.~(\ref{Keq}).
Thus, it is convenient to denote  $f(x,b,Q)= F(b) f^{\rm evol}(x,b,Q) $ and present 
the results for the the {\em evolution-generated} part $f^{\rm evol}$. 

\begin{figure}[tb]
\begin{center}
\epsfig{figure=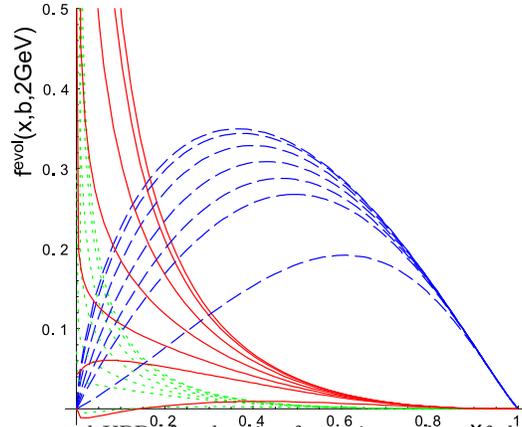,width=7cm}
\vspace{-7mm}
\end{center}
\caption{Evolution-generated UPDs for the pion for various values of the transverse
coordinate (from top to bottom: $b=0$, $1$, $2$, $3$, $4$, $5$ and
$10~{\rm GeV}^{-1}$), plotted as functions of the Bjorken $x$.  The evolution is
made with the initial condition (\ref{sqmb}) at $Q_0=313$~MeV up to
$Q=2~{\rm GeV}$.  Solid lines -- gluons, dashed lines -- valence
quarks, dotted lines -- sea quarks.}
\label{Q2}
\end{figure}

\begin{figure}[tb]
\begin{center}
\epsfig{figure=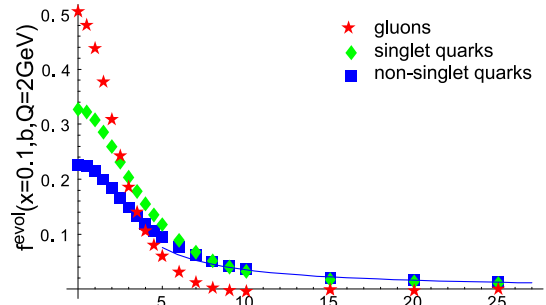,width=7cm}
\vspace{-7mm}
\end{center}
\caption{Evolution-generated UPDs for the pion for $Q=2$~GeV
and $x=0.1$, plotted as functions of $b$. The evolution is made with
the initial condition (\ref{sqmb}) at $Q_0=313$~MeV.  The numerical
results are represented by squares for the non-singlet quarks, diamonds for
the singlet quarks, and stars for the gluons, while the solid line shows the
asymptotic power-law formula for the case of non-singlet quarks. We note the 
much faster fall-of for the gluons than for the quarks.
As $Q$ is increased or $x$ decreased, the distributions in $b$ become narrower, 
which is equivalent to 
spreading in $k_\perp$.}
\label{fig:b}
\end{figure}

\begin{figure}[tb]
\begin{center}
\epsfig{figure=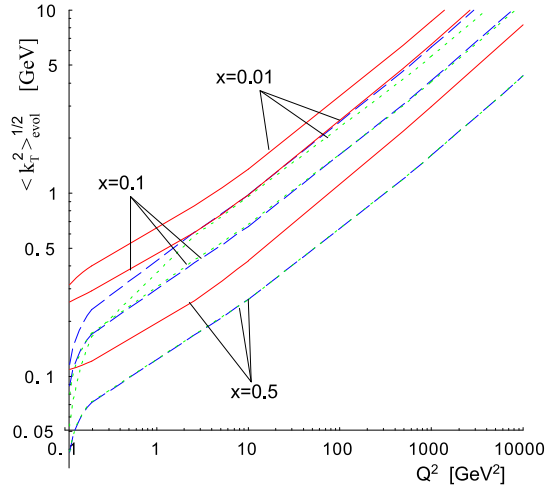,width=7cm}
\end{center}
\caption{ The rms transverse momenta of UPDs of the pion for
$x=0.01$, $0.1$, and $0.5$, plotted as functions of the
renormalization scale $Q^2$. Solid lines -- gluons, dashed lines --
non-singlet quarks, dotted lines -- singlet quarks.}
\label{fig:ktx}
\end{figure}

Equations (\ref{Keq}) can be solved very efficiently using the Mellin transform \cite{ERAWB}.
In the moment space we find
\begin{eqnarray}
Q^2 \frac{ d f_{\rm NS}({n}, {b}, Q )}{ dQ^2 } = -\frac{\alpha(Q^2)}{8 \pi}
\gamma_{{n},NS}(Q {b} ) f_{\rm NS}({n}, {b},Q), \label{seq}
\end{eqnarray} 
where $f_{\rm NS}({n}, {b}, Q )$ denote the $n$th moment, and 
$\gamma_{n,NS}(b)$ are the $b$-dependent anomalous dimensions \cite{ERAWB}.
The formal solution of Eq.~(\ref{seq}) is 
\begin{eqnarray}
{\frac{ f_{\rm NS}(n, b, Q)}{f_{\rm NS}(n, b, Q_0) } = {\rm exp} \left[
-\int_{Q_0^2}^{Q^2} \frac{d{Q'}^2 \alpha({Q'}^2)}{8\pi {Q'}^2}
\gamma_{\rm NS}(n, b, Q') \right]},
\end{eqnarray}
and similarly in the singlet channel.
In ordert to come back to the $x$-space, the inverse Mellin transform is carried out numerically.
The numerical results are shown in Figs.~\ref{Q2} and \ref{fig:b}. 
In Fig.~\ref{fig:ktx} we plot the mean squared transverse momentum generated by the evolution. 
As $Q$ is increased, $\langle k_\perp^2 \rangle$ grows as $\alpha_S(Q^2) Q^2$. 

To summarize, we have proposed an efficient numerical method to solve Eq.~(\ref{Keq})
via the Mellin transform.
The method allows to study formal aspects of the equations, {\em e.g.} the 
asymptotic forms of the evolution-generated UPDs at large $b$, or at $x\to 0$ and $x \to 1$.
At large $b$ the evolution-generated $b$-dependent UPDs exhibit power-law fall-off, with the 
magnitude of the exponents 
growing with the probing scale \cite{ERAWB}. The fall-off is steeper for the gluons than for the quarks.
At $x \to 0$ we have found generalizations of the DLLA behavior. 
We have also shown that for large $b$ the solution for the valence UPD of the pion
grows linearly with $x$ for not too large $x$, and the slope decreases with $b$ as a power law. 
At $x \to 1$ the evolution-generated $b$-dependent UPDs approach the integrated distributions as $(1-x)^2$.
We find the spreading of
the $k_\perp$ distributions with the probing scale $Q$, with the
effect strongest for gluons and increasing with decreasing $x$. We
have also shown that the widths $\langle k_\perp^2 \rangle_i^{\rm evol}$ in all
channels $i$ increase at large $Q^2$ as $Q^2 \alpha(Q^2)$.

\section*{Acknowledgements} Support
from DGI and FEDER funds, under contract BFM2002-03218 and by the
Junta de Andaluc\'\i a grant no. FM-225 and EURIDICE grant number
HPRN-CT-2003-00311 is acknowledged.  Partial support from the Spanish
Ministerio de Asuntos Exteriores and the Polish State Committee for
Scientific Research, grant number 07/2001-2002 is also gratefully
acknowledged.

\end{document}